# *Ab initio* investigation of optical properties of high-pressure phases of ice

Renjun Xu[a], Zhiming Liu, Yanming Ma, Tian Cui[b], Bingbing Liu, and Guangtian Zou

National Laboratory of Superhard Materials, Jilin University, Changchun, 130012, P. R. China

We report a detailed *ab initio* investigation on the optical properties of ice under a wide high pressure range. The ice X phase (up to 380GPa), the theoretical proposed higher pressure phase ice XV (300GPa), as well as the ambient pressure low-temperature phase ice XI are involved. Our results show that the dispersion relations of optical properties of ice under high pressure are quite different from those under ambient pressure. Under higher pressure, there is whole tendency of blue shift in all optical properties of ice, and the energy region for optical response become broader, such as the absorption band and reflection band. In addition to the augmented absorption edge, all absorption peaks are found to be rising, and the reflection peaks are also enhanced a bit; hence the transmissivity of ice is inferred to be dropping. The photoconductivity is enhanced, and we explain this behavior from the increase of earner density. The static optical properties are found to be pressure independent, and principally due to the network topology of hydrogen bonding.

[a] Current address: Department of Physics, University of California, Davis, California 95616, USA

[b] Author to whom correspondence should be addressed. Electronic address: cuitian@jlu.edu.cn

71.20.-b, 78.20.Ci, 61.50.Ks, 71.15.Mb

## I. INTRODUCTION

The optical properties for the high-pressure phases of ice have been rarely reported so far, despite its significant importance in illuminating the fundamental physical properties of ice and potential applications. In fact, 10% of the total land area is covered with glaciers at present while 32% during the last Ice Age. Understanding the interaction of sunlight with ice is important for the world climatology and satellite remote sensing [Ref. 1 2 3]. Comparing the band structures of the two high-pressure phases of ice (X and XV) with the ambient pressure phase (ice XI) in Fig. 1 [Ref. 4], we can find the maxima of their valence bands are not so flat as ice XI, indicating that they have quite different effective masses. Furthermore, the change of point group due to the high pressure phase transition should lead to different selection rules. This effect together with the altered density and band hybridization even may bring some unusual optical properties to ice. Moreover, the pressure dependence of optical properties could be used to monitor phase transitions, especially the metallization of ice [Ref. 5], which was predicted to be happen in super high pressure. As what we have analyzed in Ref. 4, the phase transitions from ice VII/VIII to ice X are induced by soft phonon modes, thus the evaluation of pressure dependence of optical properties corresponding to electronic intraband transitions is sufficient to monitor such phase transition; however, the phase transition from ice X to ice XV is predicted to be induced by the change of *s-p* charge transfers between hydrogen and oxygen, i.e. mainly the change of electronic structure, thus we should now inspect the optical properties of ice contributed by the electronic interband transitions under higher pressure. Besides, with distinct hydrogen bonding structure in ice XV (half hydrogen bonds are symmetric and the other half are asymmetric), we want to know whether there is any unusual behavior in its dielectric response.

The optical properties now known about ice under high pressure (>*60*GPa) are the infrared reflectance and absorbance [Ref[6,7,8]], and very recently the real part of refractive index[Ref.[9]]. However, all these experiments only considered about the intra-band transitions happening with incident photon energy less than 2.2eV, and they are designed for discovering the change of phonon modes associated with the transition of ice X, a symmetric hydrogen-bonded phase. In this paper, we report a detailed *ab initio* investigation of optical properties of ice due to inter-band transitions under a wide high pressure range. The ice X phase [Ref. [10,11,12]], which has a bcc substructure of oxygen atoms, the theoretically predicted higher-pressure phase ice XV [Ref. [4]], which has a distorted hcp substructure of oxygen atoms, and the ambient pressure low-temperature phase ice XI [Ref. [13,14,15,16,17,18,19,20]], a quasi-hcp structure, are all involved. Here ice XV is a new structure we found after the ion relaxation of ice XIII$^M$ [21,22], and has hydrogen-bonding structures different from other phases of ice now discovered. With the increase of pressure, a whole tendency of blue shift in all optical properties of ice is found, and it is associated with broader energy region for optical response, augmented absorption edge, and enhanced photoconductivity. In addition, all absorption peaks are found to be rising, and the reflection peaks are also enhanced a bit; thus the transmissivity of ice is inferred to be dropping and a macroscopic physical effect is predicted. At 300GPa, the predicted critical pressure of phase transition from ice X to ice XV, the similarities of optical properties between the two phases are obvious, while the differences may indicate the reason for the phase transition. From the imaginary part of dielectric functions, we argue that the phase transition from ice X to ice XV should be due to the change of *s-p* charge transfers among hydrogen and oxygen atoms. The signatures in optical properties for phase transition in ice X from 75GPa to 25GPa are also shown. Furthermore, the static real part of dielectric function $\varepsilon_l(0)$, the static reflectivity $R(0)$, and the static real part of refractivity $n(0)$ of ice are found to be pressure independent, and primarily due to the topology and framework of the 3 dimensional hydrogen bonding. In fact, these static optical properties of ice are found to be irrelevant with the bond lengths, bond angles, (e.g. the sliding of atom layers or high pressure induced change of bond lengths or bond angles), the degree of electron clouds overlap, the dispersion of electronic bands, and the band gap, as long as the framework of hydrogen bonding is not changed. As biaxial crystals, the anisotropies of refractive indices of ice XV and ice XI are also briefly discussed.

## II. THEORY AND METHOD

We perform the *ab initio* pseudopotential plane wave calculations based on the density functional theory (DFT) with norm-conserving pseudopotentials by the CASTEP code [Ref. 23]. Exchange and correlation effects are treated by a generalized gradient approximation (GGA) with the Perdew-Burke-Eruzerhof (PBE) functionals [Ref. 24] A 550eV cutoff of the kinetic energy and a self-consistent field tolerance of $5.0 \times 10^{-7}$ eV/atom are used. Moreover, we applied a finite basis set correction for the evaluation of energy and stress. The **k**-point grids for Brillouin zone sampling are generated via the

Monkhorst-Pack scheme [Ref. 25]. For different phases under different pressures, the separation of **k** points in the reciprocal space is chosen approximately 0.035 Å$^{-1}$. The crystal structures of the three phases of ice (X, XI, XV) are from Ref. 4.

Dielectric function $\varepsilon(\omega)$, which describes the features of linear response to electromagnetic radiations of a material, governs the propagation behavior of radiation in a medium. This process is induced by the proton-electron interaction. However, it is noted that we also have to take account of the momentum of phonons for indirect transitions in solids. The imaginary part $\varepsilon_2(\omega)$ could be calculated from the momentum matrix elements between the occupied and unoccupied wave functions, ensuring the selection rules. The real part $\varepsilon_1(\omega)$ could be derived from $\varepsilon_2(\omega)$ by the Kramers-Kronig transformation [Ref. 26]. All of the other optical parameters could be derived from $\varepsilon_1(\omega)$ and $\varepsilon_2(\omega)$ [Ref. 29], such as reflectivity $R(\omega)$, absorption coefficient $\alpha(\omega)$, the real and imaginary parts of the refractive index $n(\omega)+i\kappa(\omega)$, energy loss function $L(\omega)$, and optical conductivity $\sigma(\omega)$

$$R(\omega) = \left| \frac{\sqrt{\varepsilon_1(\omega)+j\varepsilon_2(\omega)}-1}{\sqrt{\varepsilon_1(\omega)+j\varepsilon_2(\omega)}+1} \right|^2,$$

$$\alpha(\omega) = \sqrt{2}\omega\left[\sqrt{\varepsilon_1^2(\omega)+\varepsilon_2^2(\omega)}-\varepsilon_1(\omega)\right]^{1/2} = 2\varpi\kappa(\varpi)/c,$$

$$n(\omega) = \left[\sqrt{\varepsilon_1^2(\omega)+\varepsilon_2^2(\omega)}+\varepsilon_1(\omega)\right]^{1/2}/\sqrt{2},$$

$$\kappa(\omega) = c\left[\sqrt{\varepsilon_1^2(\omega)+\varepsilon_2^2(\omega)}-\varepsilon_1(\omega)\right]^{1/2}/\sqrt{2},$$

$$L(\omega) = \mathrm{Im}\left[\frac{-1}{\varepsilon(\omega)}\right] = \varepsilon_2(\omega)/\left[\varepsilon_1^2(\omega)+\varepsilon_2^2(\omega)\right],$$

$$\sigma(\omega) = \sigma_1(\omega) + i\sigma_2(\omega) = -i\omega\varepsilon_0[\varepsilon(\omega) - 1].$$

The reliability and accuracy of the calculation method we use has been confirmed by many recent papers, e.g. in Refs. 27, 28, and 29. Because of the limited number of empty bands and *k*-points we can use due to the cost of computation, the complete fine structures could not be shown in our figures. Nevertheless, we use the Gaussian broadening method for the smearing of our curves, thus most peaks can be identified. Because the electron energy loss function in the high energy region is greatly affected by the number of empty bands, we use 36 empty bands to produce reasonable optical properties which cover a wide energy range. Since ice X is optically isotropic while ice XI and ice XV are optically anisotropic, we will firstly discuss the optical properties of the three phases as polycrystals, in which the optical properties are averaged over all possible polarization directions (this makes no difference for ice X because of its isotropy).

## III. RESULTS AND DISCUSSIONS

We plot the imaginary parts of the dielectric functions $\varepsilon_2(\omega)$ for ice X (under 25GPa, 75GPa, and 300GPa), ice XV (under 300GPa), and ice XI (under ambient pressure), in Figs. 2a-c, respectively. The peaks in the low energy region of $\varepsilon_2(\omega)$ represent the intrinsic optical absorption, and reflect the electronic structure near Fermi surface; the high energy region represent the characteristic spectrum of a material, and is due to the joint density of states between the valence and conduction bands. The partial density of states for ice X, ice XV, and ice XI are presented in Fig. 3 [Ref.[4]].

In Fig. 2a for $\varepsilon_2(\omega)$ of ice X, from 25GPa to 300GPa, there is a whole tendency of blue shift (move toward higher energy region) due to the increase of the density of electrons. In fact, this pressure-induced blue shift effect is also reflected in all of the other optical properties as is shown in other figures below. Furthermore, one peak at 18.5eV under 25GPa disappears when ice X is compressed to 75GPa, indicating there must be a phase transition for ice X between 25-75GPa. This phase transition has been confirmed by many experiments [Ref. [30][31]]. Now we explain the origin of every major peak of ice X under 300GPa here for illustration. The two most prominent peaks at 15.4eV and 16.5eV above the optical absorption edge are dominantly contributed by the interband transitions from the highest occupied states (mainly the lone pair *p* states from oxygen corresponding to the two peaks of DOS between -2.4eV and -1.3eV) to the lowest unoccupied states with electronic energy 13.9eV, which correspond to an anti-bonding orbital from *s* states of oxygen and hydrogen. The peak at 21.2eV is related to the transition also from the highest occupied states but to the secondary lowest unoccupied states at 18.9eV, which belong to *s-p* hybridized states.

Comparing the dispersion curve of $\varepsilon_2(\omega)$ of ice XV (Fig. 2b) with ice X both under 300GPa, we can find that they have a similar line shape, especially the positions of the two peaks, respectively at 16.0eV and 21.2eV. The physical processes for them in ice XV are very similar to those in ice X. However, we could see that the second major peak at 21.2eV is much weaker than that in ice X. This peak corresponding to a transition excited to an *s-p* hybridized unoccupied state, hence we suspect that the phase transition from ice X to ice XV is very likely induced by the *s-p* charge transfers.

Turning to Fig.2c, we can find the dispersion curve $\varepsilon_2(\omega)$ of the ambient pressure phase of ice (ice XI) is very different from both of the two high-pressure phases (ice X and ice XV). There are more complex structures, especially in the comparatively low photon frequency region between 5.2eV and 14.8eV, which are contributed by the interband transitions from the bands corresponding to four molecular energy levels ($\pi_y^{non}, \sigma_z^*, \sigma_z$, and $\sigma_x^{non}$) to the unoccupied states below 14.4eV. The other four prominent peaks in the high photon frequency region at 15.6eV, 17.4eV, 18.5eV, and 21.4eV are contributed by the transitions respectively from the bands corresponding to the four molecular energy levels mentioned above to the CBs in the high DOS region around the DOS peak at 15.4eV.

The real parts of dielectric function $\varepsilon_1(\omega)$ for the three phases are plotted in Fig. 4. Comparing $\varepsilon_1(\omega)$ of ice X under 75GPa with300GPa, we can also find the whole tendency of blue shift, but the static real part of the dielectric constant, $\varepsilon_1(0)$, is nearly pressure independent, only dropped from 3.80 at 75GPa to 3.72 at 300GPa. One peak at 18.0eV under 25GPa disappears when ice X is compressed to 75GPa. Furthermore, we note that the average value of $\varepsilon_1(0)$ are nearly the same between ice X and ice XV, i.e. about 3.78, which are nearly the double value of that in ice XI (1.98). These behaviors indicate that the key factor affecting the value of $\varepsilon_1(0)$ in ice is the topological structure and framework of the 3 dimensional hydrogen bonding; Within the same framework, $\varepsilon_1(0)$ is irrelevant with the bond lengths and bond angles (e.g. the sliding of atom layers or high pressure induced changes of them), the degree of electron clouds overlap, the dispersion of electronic bands, and the band gap. It is worthy to point out that with such high dielectric constants, ice should conduct electricity much better than most nonmetallic crystals. Hence, in order to find a metallic phase of ice, we must search topologies and frameworks of hydrogen bonding which are different from any known phases of ice; neither ice X nor ice XV could be metallized under the pressure we investigated or even higher.

There are two key optical parameters for the remote measurement of ice conditions, namely the absorption and reflection coefficients, because when a beam passes through a piece of ice the decrease in the intensity is determined by both the absorption and the scattering of energy. The physical scheme happened in this process can be described by the equation

$$I_x(\omega) = [1 - R(\omega)]I_0(\omega)\exp[-\alpha(\omega)x],$$

where $I_x(\omega)$ is the intensity at distance $x$ beneath the surface, and $\alpha(\omega)$ is defined as the absorption coefficient. Fig. 5 shows the absorption coefficients $\alpha(\omega)$ for the three phases. In Fig. 5a for ice X from 25GPa to 75GPa, the second major peak at 16.6eV disappears; from 75GPa to 300GPa, accompanying the blue shift of the absorption spectra, the optical absorption edges are also increased, and the corresponding values are in good agreement with the band gaps of ice [Ref.[4]] (any photon having energy within the forbidden band is just transparent to a material). Moreover we find that all absorption peaks are raised. Hence there is a pressure-induced physical effect: the ice will turn to be more opaque under higher pressure for the photon energy much greater than the optical absorption edge — due to the constancy of the energy stream, the transmissivity must be dropped. Furthermore, we find that the absorption bands become wider under higher pressure, i.e. from (7.0eV – 100.5eV) to (10.5eV – 120.5eV). The ice XV is similar to ice X in the profile of $\alpha(\omega)$ spectra, but slightly different in weaker absorption peaks and narrower absorption bands. Comparing with the two high-pressure phases, ice XI is quite different in that there are three prominent absorption peaks and even narrower optical responsive energy range due to its band structure. In addition, the absorption of all the three phases decreases with the photon energy in the high energy region, where the electron is hard to respond. Take ice X under 300GPa for example, the absorption comes to be very weak for photon energy greater than 56.0eV, is due to the energy limitation of the highest unoccupied states of H *1s*, O *2s*, and O *2p*, which could be seen from the spectra of partial density of states of ice X in Fig. 3. Hence, to make excitation corresponding to the higher photon energy possible, the hydrogen and/or oxygen have to be excited to H *2s* and/or O *3s* states, and this process is more difficult to happen.

In Fig. 6a for the reflective spectra of ice X from 25GPa to 75GPa, the second major peak at 16.8eV disappears, indicating a signature of phase transition. Comparing reflective spectra of ice X and ice XV both under 300GPa in Fig. 6, we find that both of their reflectivities decrease dramatically when the photon energy increases from 30.5eV to 50.0eV, indicating there should be strong peaks appear in the electron energy loss spectrum (EELS) in this photon energy range, the trailing edges of the reflection spectra. In addition, we note that they have nearly the same value of *R(0)*, ~0.10, which is triple that of ice XI, 0.03. The difference is that ice XV has smaller reflection bands (0-93.6eV) than ice X (0-106.8eV), and both of them are much larger than that of ice XI (0-41.4eV).

In Fig. 7, we present the electron energy loss function $L(\omega)$, an important parameter describing the energy loss of a beam of fast electrons traversing in a material. Calculating $\frac{dL(\omega_p)}{d\omega} = 0$ from the formula of $L(\omega)$ given in this paper, we can get $\varepsilon_1(\omega_p) = 0$, thus the peaks in $L(\omega)$ spectra represent the plasma resonance frequencies, above which the material behaves dielectric [$\varepsilon_1(\omega) > 0$] while below which the materials behaves metallic [$\varepsilon_1(\omega) < 0$]. There is a broad peak appears between ~30.5eV and ~50eV respectively for ice X and ice XV under 300GPa. This is in good agreement with the prediction from Fig. 6. The EELS of ice XI under ambient pressure is quite different from them in that there is one prominent peak at ~22.1eV, and another weaker one at ~16.3eV, both of which are the plasma frequencies corresponding to two metallic regions in $\varepsilon_1(\omega)$ spectrum where $\varepsilon_1(\omega) < 0$. The EELS signature of phase transition for ice X from 25GPa to 75GPa is the disappearance of one major peak at 28.0eV.

Photoconductivity is the increase in electrical conductivity caused by the excitation of additional free charge carriers by photons with sufficiently high energy beyond the forbidden band gap. Since the electrical conductivity of a material is given by the product of the carrier density, its charge, and its mobility, an increase in the photoconductivity in single-crystal materials should be due primarily to an increase in earner density. This accounts for the increase of conductivity in X from 75GPa to 300GPa (Fig. 8), e.g. the strongest peak increased from 8.0 to 11.0, at 17.7eV and 21.2eV, respectively. The signature of phase transition for ice X from 25GPa to 75GPa is the disappearance of one major peak at 18.6eV. At present, we do not know what the photoexcited charge carriers are for sure. We suppose that they might be protons due to some kind of light induced defects, or $H_3O^+$ ions and negative complex trapping electrons. Comparing the photoconductivities among the three phases, we can find that the two high-pressure phases have similar dispersions, but both are quite different from that in the ambient pressure phase ice XI. Hence there must be heterogeneous photoexcitation process and scheme of photoconductivity in ice under high pressure. To understand the whole process completely, further theoretical and experimental research need to be carried out.

Corresponding to $\varepsilon_1(\omega)$ and $\varepsilon_2(\omega)$, the dispersion relations of the real and imaginary parts of refractive indices, $n(\omega)$ and $k(\omega)$, of ice X are seldom changed by pressure except the whole tendency of blue shift, as illustrated in Fig. 9. As same as the relationship between $\varepsilon_1(\omega)$ and $\varepsilon_2(\omega)$, the maximum and minimum values of $n(\omega)$ correspond to the photon frequencies where $k(\omega)$ have maximum gradients in raising and in dropping, respectively. Comparing the three phases, we can find: i. The static real parts of the refractive indices, $n(0)$, are the same between ice X (75GPa-300GPa) and ice XV (300GPa), 1.9, which is larger than that of ice XI under ambient pressure, 1.4; ii. $n(\omega)$ of both the two high-pressure phases decrease sharply above the photon energy corresponding to the second major peak of $\varepsilon_1(\omega)$ and below the plasma frequency, and this behavior is quite different from the low-pressure phase ice XI, which has comparatively complicated dispersion. The signature of phase transition for ice X from 25GPa to 75GPa is the disappearance of peaks at 18.2eV in $n(\omega)$ and at 19.3eV in $k(\omega)$.

Since ice XV and ice XI are optically biaxial crystals, we present the real parts of the refractive indices respectively along the [100], [011], and [111] polarization directions for them in Fig. 10, where we could see the significant differences among $n_{[100]}$, $n_{[011]}$, and $n_{[111]}$ respectively in the two phases.

## IV. CONCLUSION

We carried out detailed investigation on the optical properties fro the two high-pressure phases of ice (ice X and ice XV), and compared them with the ambient pressure phase ice XI using the *ab initio* pseudopotential density functional method. When ice is compressed, there is whole tendency of blue shift in all optical properties of ice, and the optical response energy region becomes broader. This behavior is due to the blue shift of conduction bands and the red shift of valence bands, and thus the larger band gap. The augmented optical absorption edge indicates that ice X should behave dielectrically better under higher pressure, but not metallically. Besides, we found that all absorption peaks are raising, and the reflection peaks are also enhanced a bit, thus the transmissivity must be dropped. Hence a pressure-induced macroscopic effect is predicted: the ice will turn to be more opaque under higher pressure for the photon energy much greater than the optical absorption edge. The photoconductivity of ice X is also enhanced, and primarily due to the increase of earner density; however, the microscopic physical picture about the photoexcitation process under high pressure is still unclear, and need further research. The signatures in optical properties for phase transition in ice X from 75GPa to 25GPa are presented. Under the critical phase transition pressure 300GPa, though ice XV and ice X have some similar optical properties, they have differences. For instance, from the imaginary part of dielectric functions of them, we argued that the phase transition from ice X to ice XV should be due to the *s-p* charge transfers among hydrogen and oxygen atoms. Furthermore, we find some common behavior in optical properties of ice under different pressures. The static optical properties of ice, e.g. $\varepsilon_1(0)$, $R(0)$, and $n(0)$, are found to be pressure independent. The key factor that affects these values is the topology and framework of the 3 dimensional hydrogen bonding, and they are irrelevant with the bond lengths, bond angles, (e.g. the sliding of atom layers or high pressure induced change of bond lengths or bond angles), the degree of electron clouds overlap, the dispersion of electronic bands, and the band gap, as long as the framework of hydrogen bonding is not changed. Hence, in order to find a metallic phase of ice, we must search hydrogen bonding network structures different from any known phases of ice; neither ice X nor ice XV could be metalized under the pressure we investigated or even higher.

## ACKNOWLEDGEMENTS

This work was supported by the National Natural Science Foundation of China under grant No. 10574053 and 10674053, 2004 NCET and 2003 EYTP of MOE of China, the National Basic Research Program of China, Grant No. 2005CB724400 and 2001CB711201, and the Cultivation Fund of the Key Scientific and Technical Innovation Project, No. 2004-295.

# List of Figures

FIG. 1. (Color online) Band structures of: (a) ice X under 300GPa, (b) ice XV under 300GPa, and (c) ice XI under the ambient pressure [Ref. 4].

FIG. 2. (Color online) Imaginary part of dielectric function $\varepsilon_2(\omega)$ calculated for (a) ice X under 25GPa, 75GPa, and 300GPa, (b) polycrystalline ice XV under 300GPa, and (c) polycrystalline ice XI under ambient pressure.

FIG. 3. (Color online) Spectra of partial density of states for (a) ice XI under ambient pressure, (b) ice X under 75 GPa, (c) ice X under 300GPa, and (d) ice XV under 300GPa [Ref. 4].

FIG. 4. (Color online) Real part of dielectric function $\varepsilon_1(\omega)$ calculated for (a) ice X under 25GPa, 75GPa, and 300GPa, (b) polycrystalline ice XV under 300GPa, and (c) polycrystalline ice XI under ambient pressure.

FIG. 5. (Color online) Absorption coefficients calculated for (a) ice X under 25GPa, 75GPa, and 300GPa, (b) polycrystalline ice XV under 300GPa, and compared with (c) polycrystalline ice XI under ambient pressure.

FIG. 6. (Color online) Reflectivity of (a) ice X under 25GPa, 75GPa, and 300GPa, (b) polycrystalline ice XV under 300GPa, and (c) polycrystalline ice XI under ambient pressure.

FIG. 7. (Color online) The electron energy loss spectrum of (a) ice X under 25GPa, 75GPa, and 300GPa, (b) polycrystalline ice XV under 300GPa, and the comparison with that of (c) polycrystalline ice XI under ambient pressure.

FIG. 8. (Color online) Real part of photoconductivity of (a) ice X under 25GPa, 75GPa, and 300GPa, (b) polycrystalline ice XV under 300GPa, and (c) polycrystalline ice XI under ambient pressure.

FIG. 9. (Color online) Refractive indices of (a) ice X under 25GPa, 75GPa, and 300GPa, (b) polycrystalline ice XV under 300GPa, and (c) polycrystalline ice XI under ambient pressure.

FIG. 10. (Color online) Anisotropic refractive index $n(\omega)$ of (a) ice XI under ambient pressure and (b) ice XV under 300GPa.

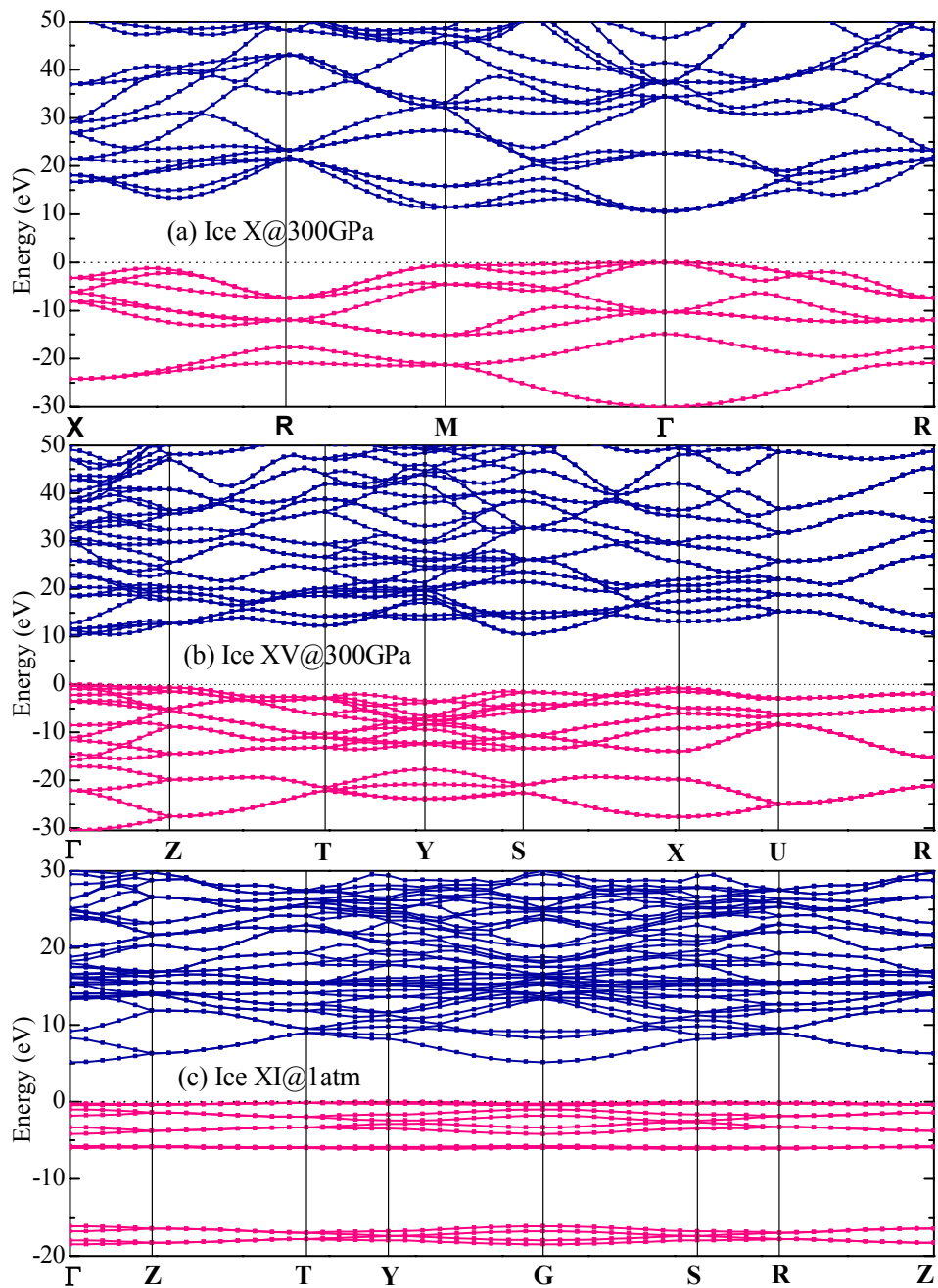

FIG. 1

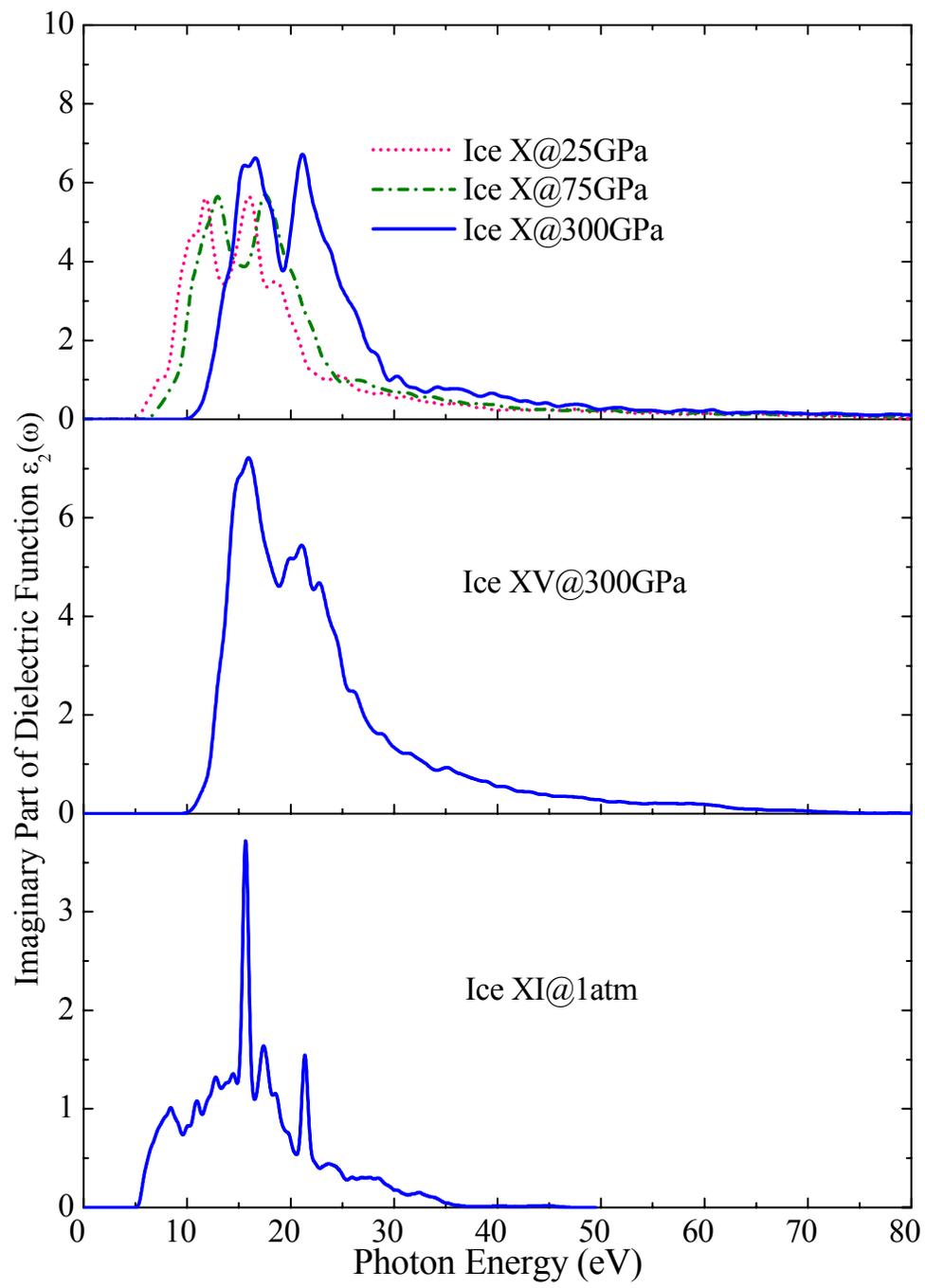

**FIG. 2**

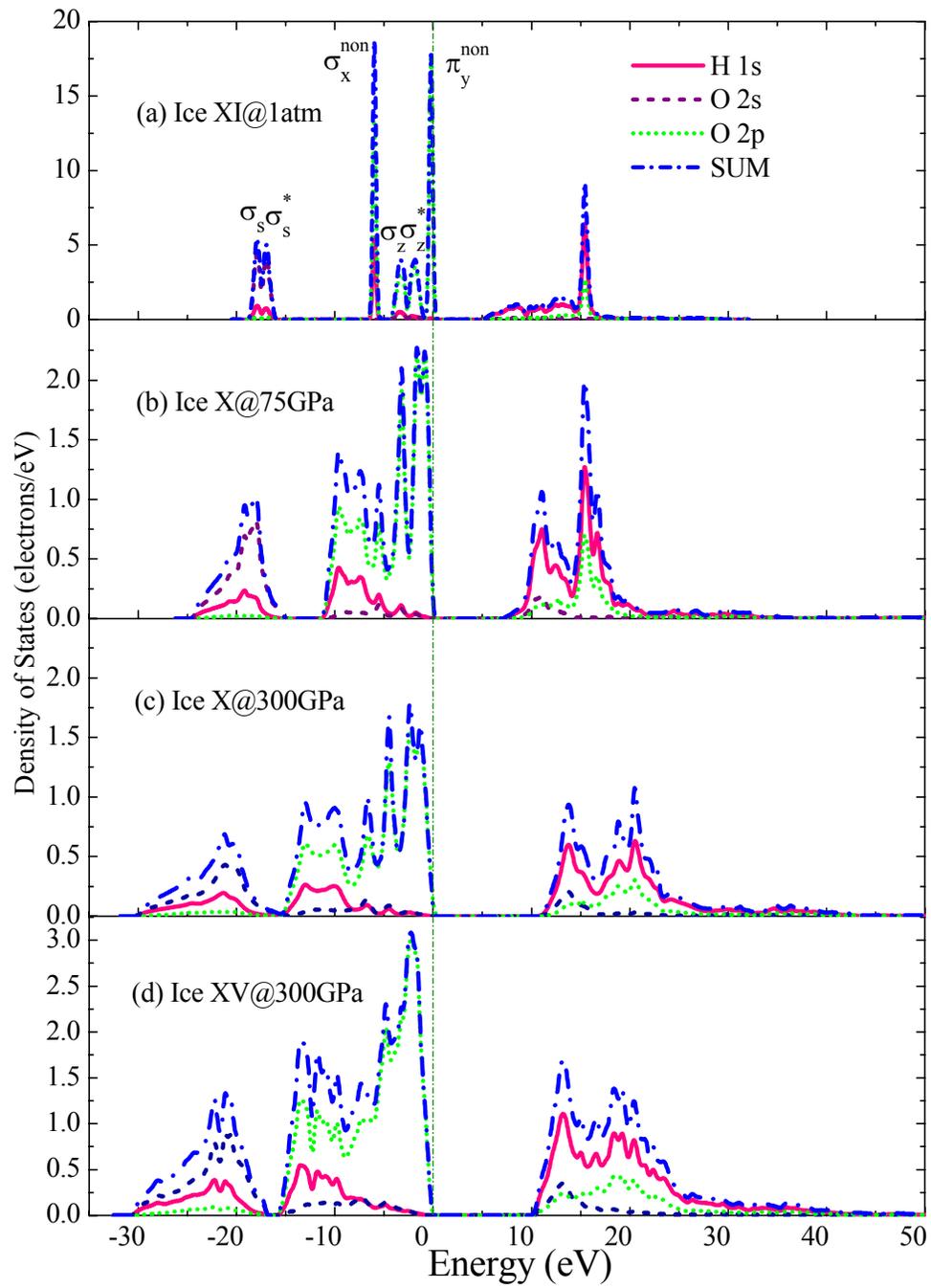

**FIG. 3**

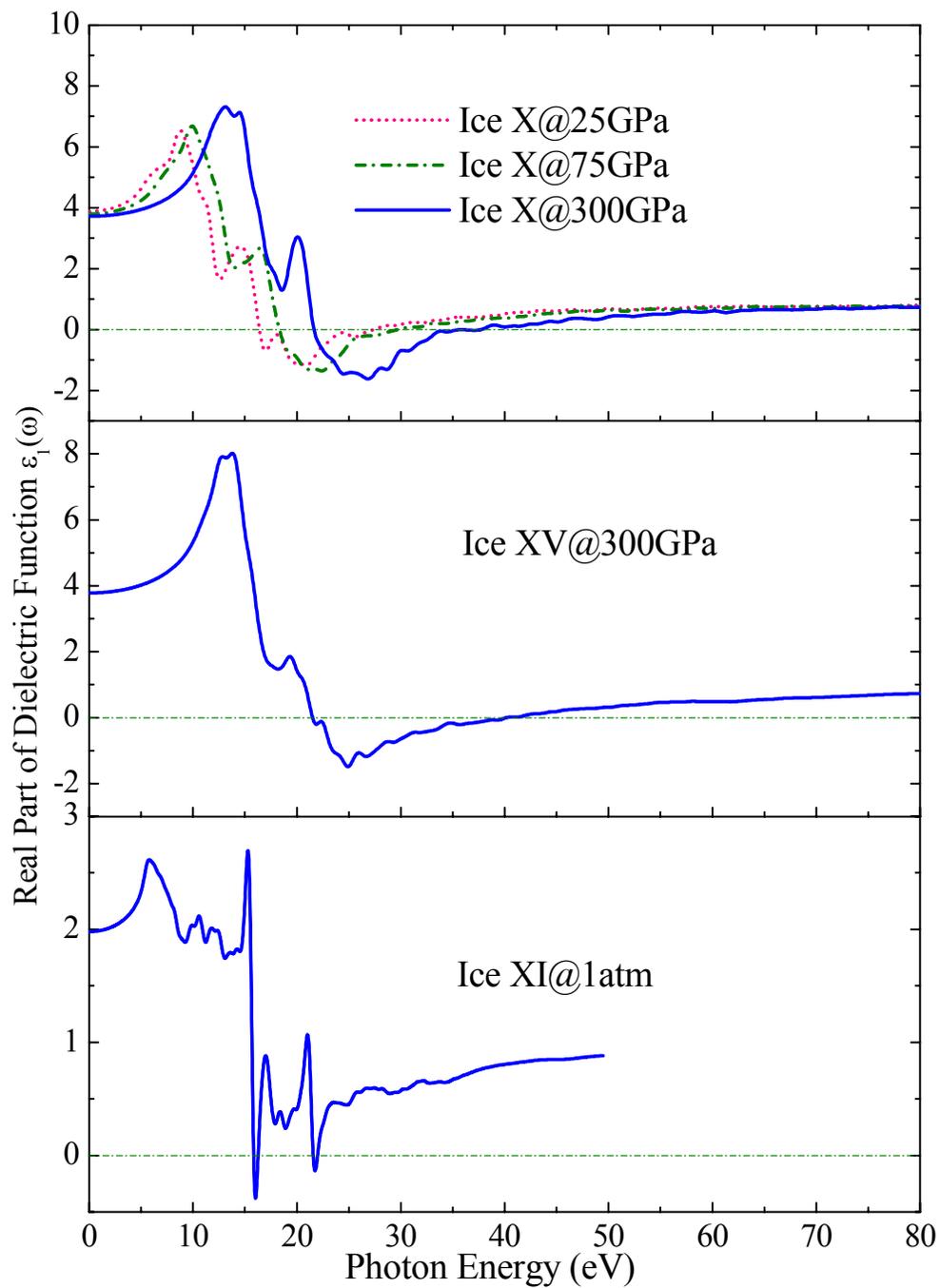

**FIG. 4**

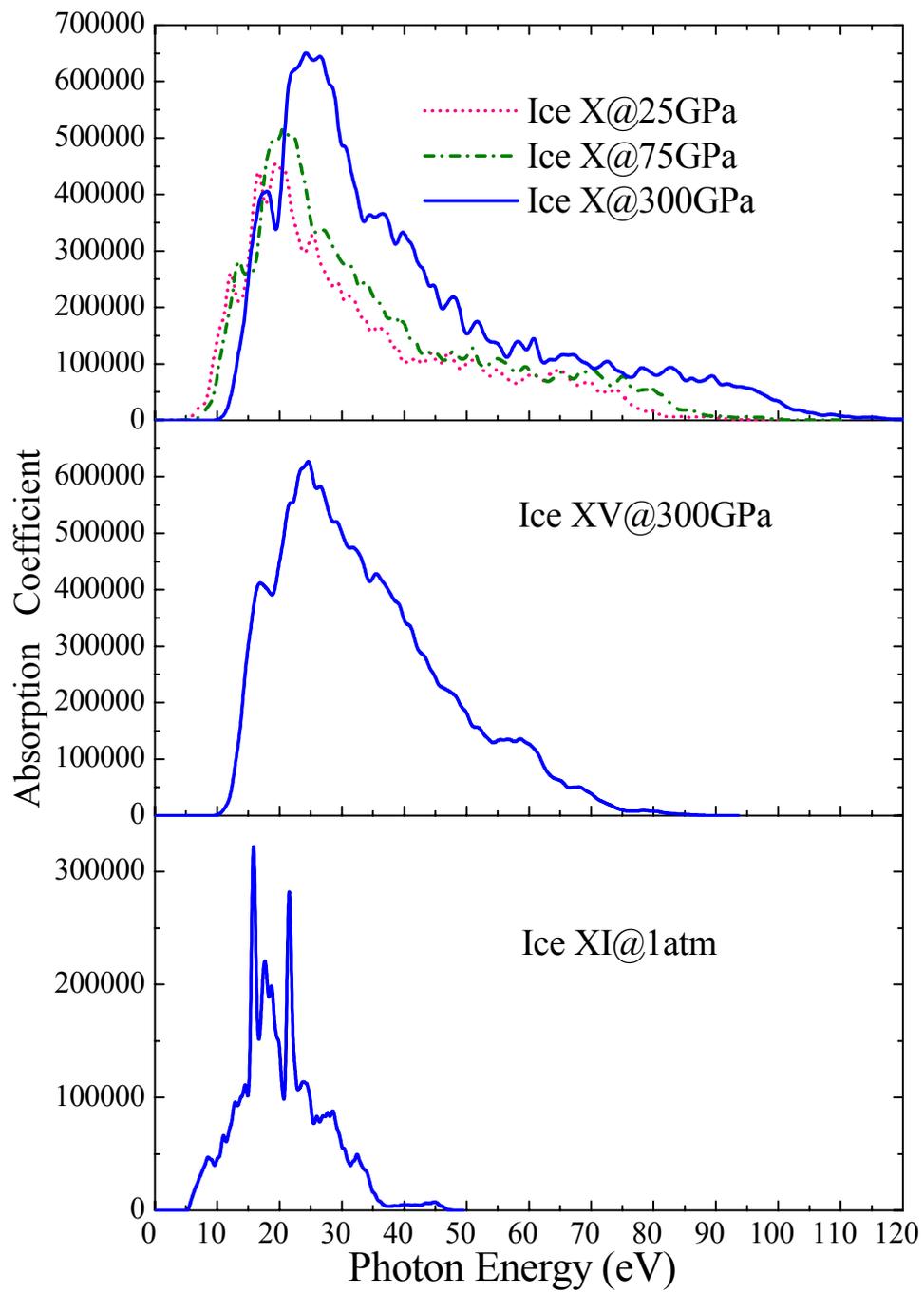

**FIG. 5**

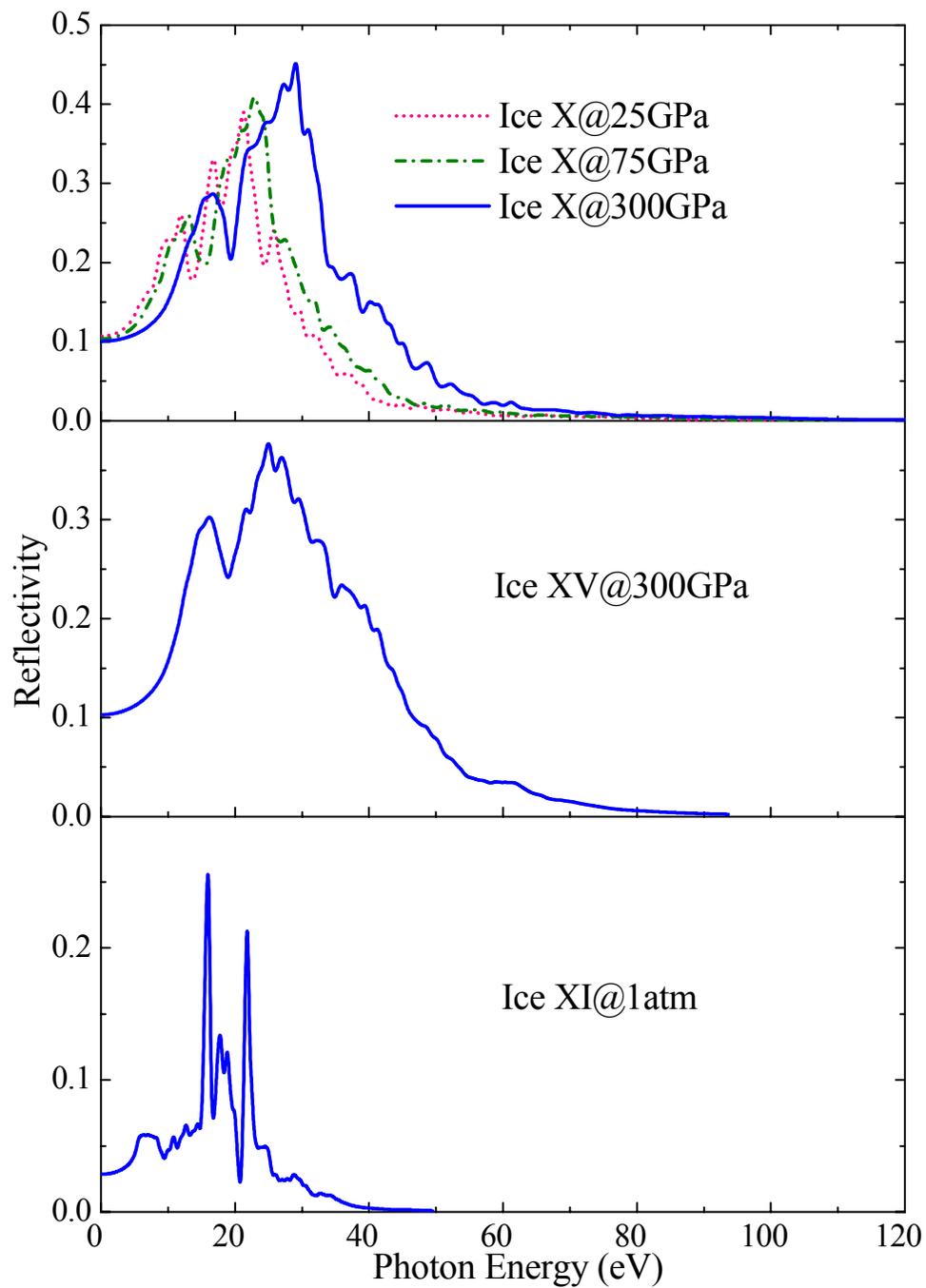

**FIG. 6**

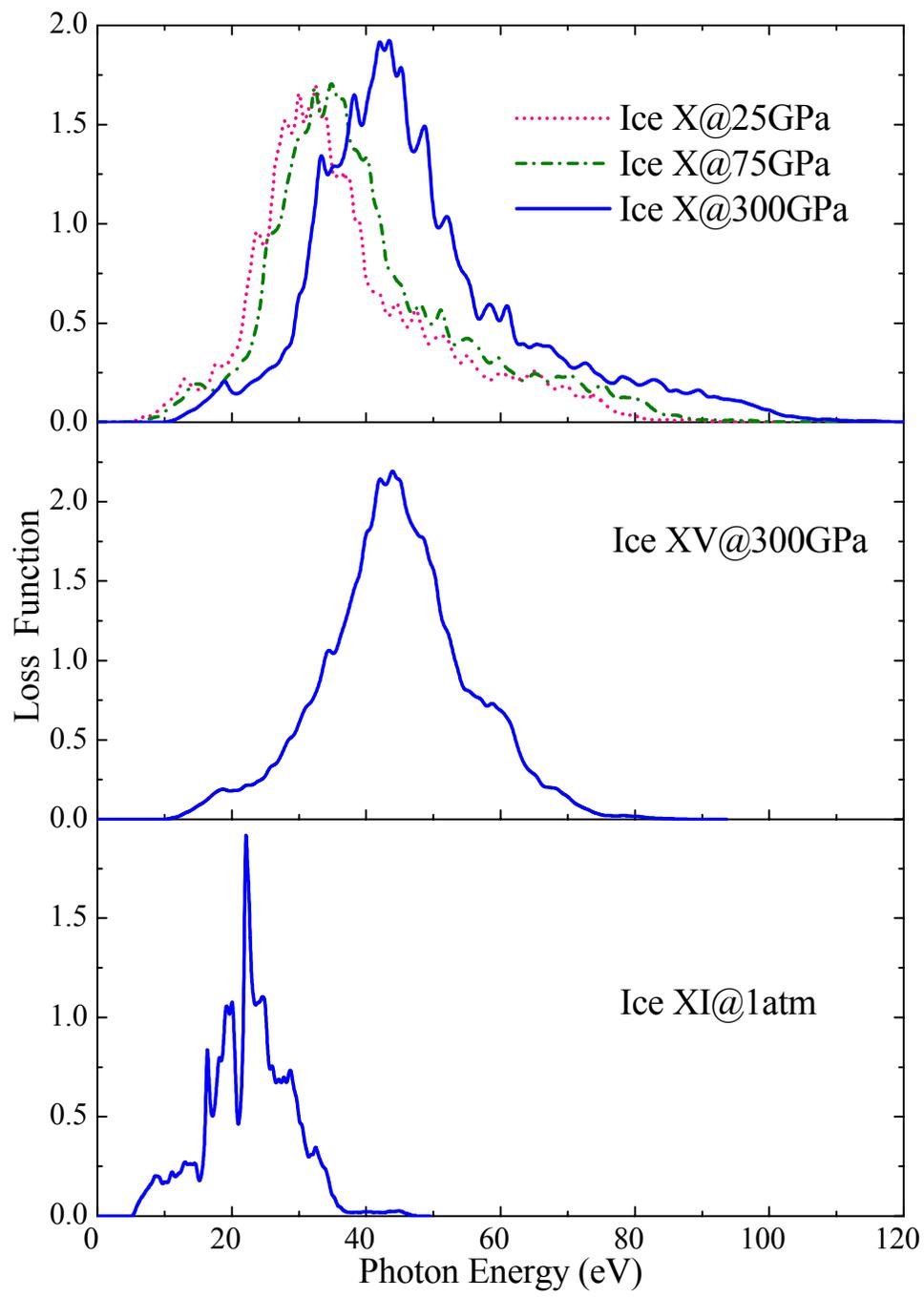

**FIG. 7**

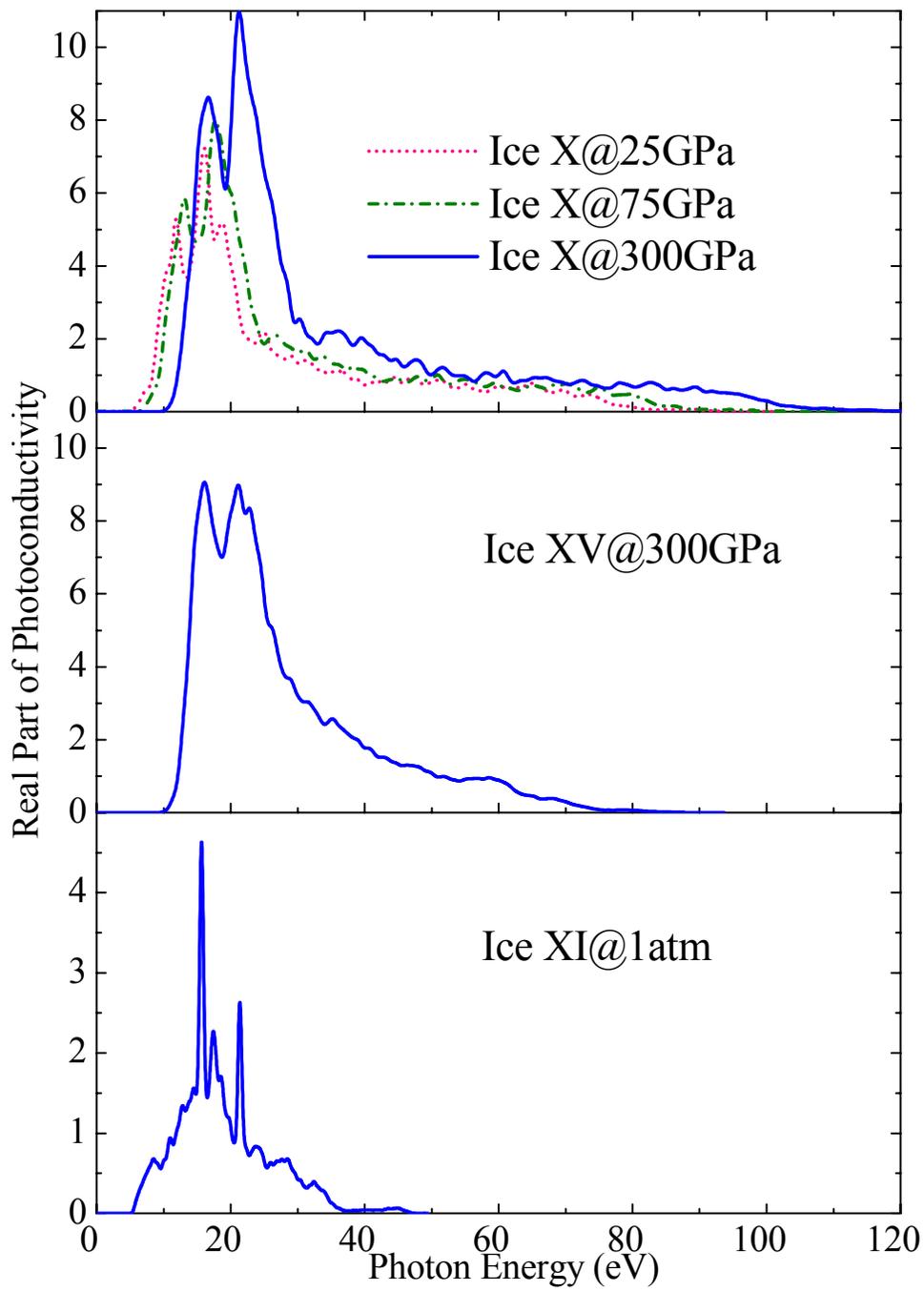

**FIG. 8**

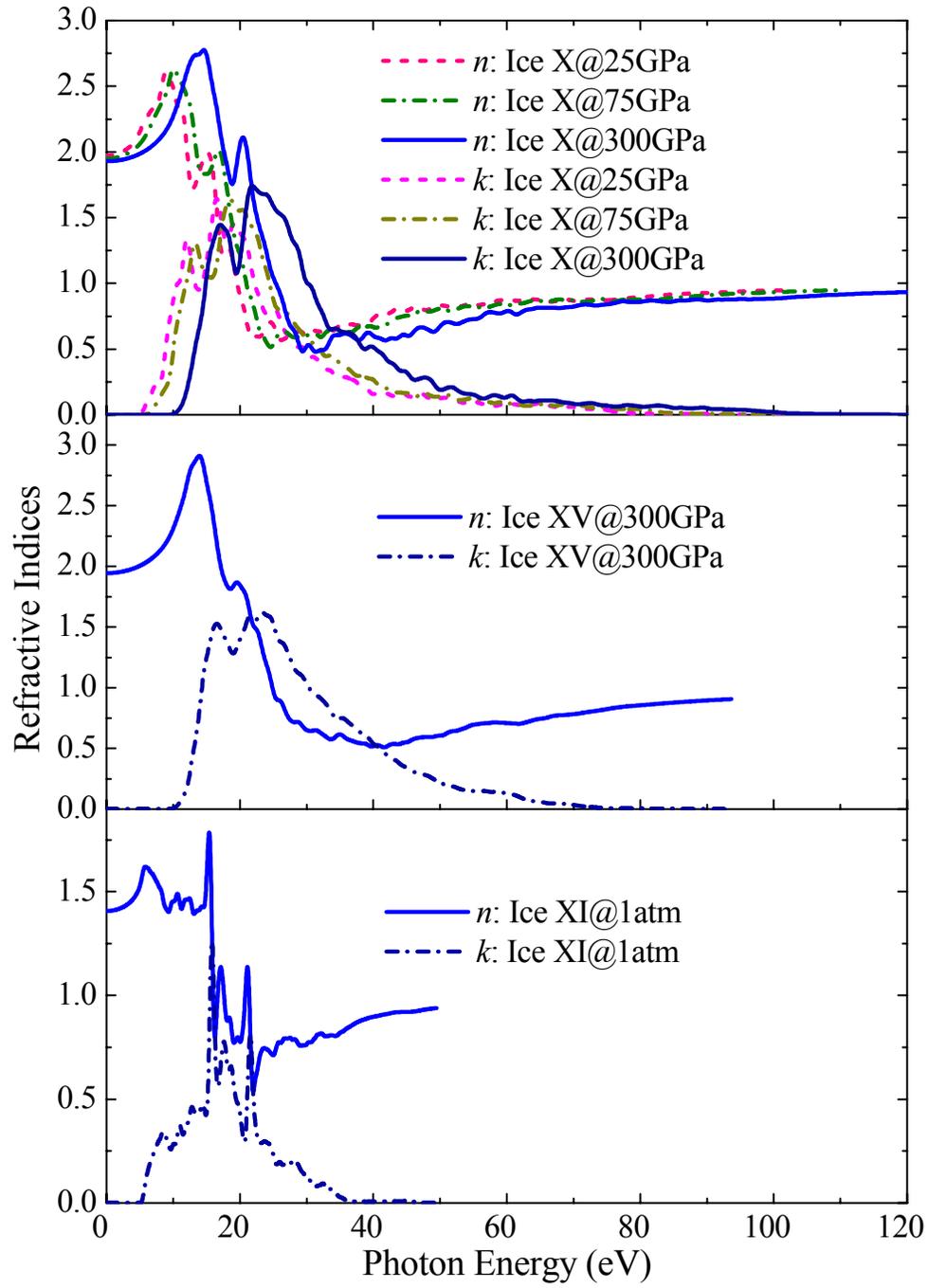

**FIG. 9**

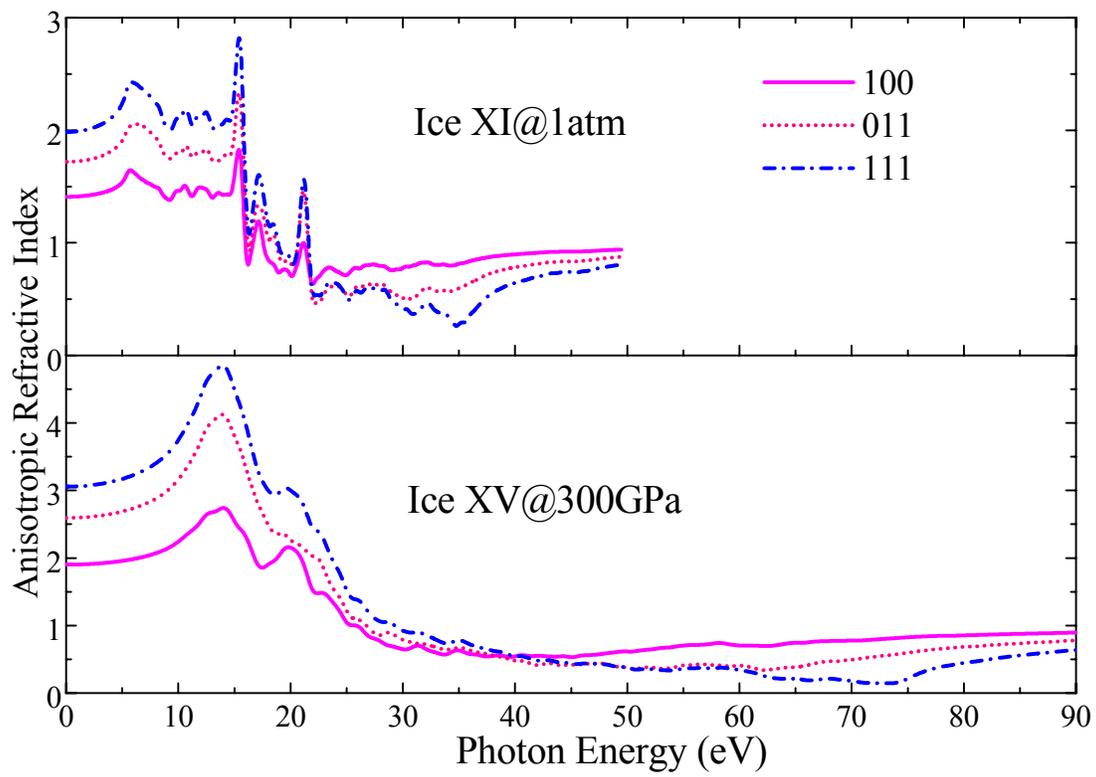

**FIG. 10**